\documentclass[amsmath,amssymb,showpacs]{revtex4}
\usepackage{amsmath,amsfonts,stmaryrd}
\usepackage{hyperref}
\begin{document}

\title{Consistent holographic description of boost-invariant plasma} 

\author{Michal P.~Heller}
\affiliation{
\it Institute of Physics, Jagiellonian University\\
\it Reymonta 4, 30-059 Cracow, Poland}
\email[]{michal.heller@uj.edu.pl}

\author{R.~Loganayagam}
\affiliation{
\it Department of Theoretical Physics, Tata Institute of Fundamental Research\\
\it Homi Bhabha Rd, Mumbai 400005, India}
\email[]{nayagam@theory.tifr.res.in}

\author{Micha\l\ Spali\'nski}
\affiliation{
\it So{\l}tan Institute for Nuclear Studies\\
\it Ho{\.z}a 69, 00-681 Warsaw, Poland}
\email[]{michal.spalinski@fuw.edu.pl}

\author{Piotr Sur{\'o}wka}
\affiliation{
\it Institute of Physics, Jagiellonian University\\
\it Reymonta 4, 30-059 Cracow, Poland}
\email[]{surowka@th.if.uj.edu.pl}

\author{Samuel E.~V{\'a}zquez}
\affiliation{
\it Perimeter Institute for Theoretical Physics\\
\it 31 Caroline St.~North,\\
\it Waterloo, ON, Canada N2L-2Y5}
\email[]{svazquez@perimeterinstitute.ca}


\newcommand{\ud}{\mathrm{d}}
\newcommand{\eq}{\begin{equation}}
\newcommand{\eqx}{\end{equation}}
\newcommand{\eqn}{\begin{eqnarray}}
\newcommand{\eqnx}{\end{eqnarray}}
\newcommand{\ttau}{\tilde{\tau}}
\newcommand{\spacer}{\vspace{6 pt}}

\begin{abstract}
Prior attempts to construct the gravity dual of boost-invariant flow of 
$\mathcal{N}=4$ supersymmetric Yang-Mills gauge theory plasma suffered from
apparent curvature singularities in the late time expansion. This Letter
shows how these problems can be resolved by a different choice of expansion
parameter. The calculations presented correctly reproduce the plasma 
energy-momentum tensor within the framework of second order viscous
hydrodynamics. 
\end{abstract}

\maketitle

\section*{Introduction}
Gauge/string theory duality \cite{adscft} has proved to be a valuable tool in 
describing the properties of strongly coupled gauge theory at finite
temperature. Most activity has focused on static
situations or linearized dynamics. Only recently some progress was made in
the studies of strongly coupled nonlinear gauge
theory dynamics using string theory methods
\cite{Janik:2005zt, Nakamura:2006ih, Janik:2006ft, Heller:2007qt, Benincasa:2007tp,
Buchel:2008xr, Buchel:2008qw, Bhattacharyya:2008jc,
Bhattacharyya:2008xc,VanRaamsdonk:2008fp,Dutta:2008gf}.  
Such studies are currently of great practical importance due to
experimental investigations of strongly coupled QCD plasma at RHIC and soon
at 
the LHC \cite{review}. While the AdS/CFT correspondence is well understood
only in the supersymmetric setting, it is believed that 
this case can capture some essential features of real-world QCD above the
deconfinement temperature. 
 
\spacer

Experimental data suggest that a boost invariant description of the
expansion of the fireball seen at RHIC should provide a useful model.  
This is based on Bjorken's observation \cite{Bjorken} that multiplicity
spectra when 
expressed in proper time and rapidity variables are approximately
independent of rapidity in the mid-rapidity region. 
The pioneering work of Janik and Peschanski \cite{Janik:2005zt} established 
the gravity dual of the boost-invariant flow of $\mathcal{N} = 4$ plasma in
the regime of large proper time.
These authors showed that asymptotic large time behavior of gauge theory
plasma, when matter is locally well
equilibrated, is given by hydrodynamics. 
The large proper time expansion is equivalent to a gradient expansion of
hydrodynamics. In a series
of follow-up papers \cite{Nakamura:2006ih, Janik:2006ft, Heller:2007qt}
sub-asymptotic corrections were calculated, corresponding to various
dissipative terms in the hydrodynamical description. These results 
were of great interest to the heavy ion community, because they provided 
numerical values of strongly coupled gauge theory transport coefficients 
starting from first principles. 
Obtaining the energy-momentum tensor of
boost-invariant flow up to second order also helped to
establish the correct theory of causal conformal hydrodynamics
\cite{Baier:2007ix}. 

\spacer

The approach explored in
\cite{Janik:2005zt,Janik:2006ft} relied
entirely on demanding 
that coefficients in the expansion of curvature invariants in powers of
inverse proper time should be regular. 
This determined the
viscosity coefficient in a way consistent with results obtained by other
means \cite{Policastro:2001yc} and for the first time provided information
about second order viscous hydrodynamics for strongly coupled plasma
\cite{Heller:2007qt}. However, this 
regularity condition turned out to be 
violated at third order in the large proper time expansion
\cite{Heller:2007qt}. 
It was suggested \cite{Benincasa:2007tp, Buchel:2008xr} that the 
singularities encountered cannot be canceled within the supergravity
approximation and they indicate either the need for
additional string theory degrees of freedom, or a genuine instability.
The
intention of this Letter is to readdress this issue.

\spacer

Recently another framework describing gauge theory plasma hydrodynamics was
developed \cite{Bhattacharyya:2008jc}, where the gravity dual is determined
order by order in a gradient expansion starting from a locally boosted
black brane geometry. This approach utilizes generalized incoming
Eddington-Finkelstein coordinates and yields a manifestly regular metric
\cite{Bhattacharyya:2008jc,Bhattacharyya:2008xc,VanRaamsdonk:2008fp,Dutta:2008gf}.
In \cite{Bhattacharyya:2008jc} the energy-momentum tensor was found
explicitly up to second order in the gradient expansion; the calculation
makes it is hard to envisage 
how singularities could arise at third order, even though such a
calculation seems technically rather demanding. It is natural to ask
whether this experience can be used to shed light on the results of
\cite{Heller:2007qt}. The answer is affirmative. A careful analysis of the
late-time boost-invariant geometry employing the incoming Eddington-Finkelstein
coordinates shows (as discussed in this Letter) that singularities
encountered in \cite{Heller:2007qt, Benincasa:2007tp} are unphysical and
arise due to the choice of expansion parameter. 
The coordinates used here
are suitable, since the corrections in the gradient expansion remain valid
up to the curvature singularity at the origin. Note that while the
curvature invariants are diffeomorphism-invariant, both their expansion
coefficients in powers of proper time and the expansion parameter itself
are not. This is the reason why a judicious choice of
coordinates is relevant for the formulation of a consistent late time expansion.

\section*{Boost-invariant flow from the black brane solution}

Bjorken expansion \cite{Bjorken} of $\mathcal{N} = 4$ SYM plasma is a
one-dimensional flow with boost invariance along the expansion axis and
rotational and translational symmetries in the perpendicular plane
\cite{Bjorken}. Proper time $\tau$ and rapidity $y$ are related to the
usual lab-frame coordinates by $x^{0} = \tau \cosh{y},\quad x^{1} = \tau
\sinh{y}$. Proper time $\tau$ is invariant under boosts along the collision
axis, whereas rapidity $y$ is not. Thus boost invariance implies that
physical quantities can depend only on proper time $\tau$, not on rapidity
$y$.

\spacer

Following the ideas of
\cite{Bhattacharyya:2008jc} 
boost-invariant perfect fluid flow can be obtained locally from the
5-dimensional boosted black brane solution
\eq
\ud s^2 = -2 u_\mu \ud x^\mu \ud r - r^2 \left(1-\frac{1}{b^4 r^4}\right)
u_\mu u_\nu \ud x^\mu \ud x^\nu + r^2 \left(\eta_{\mu\nu} + u_\mu
u_\nu\right) \ud x^\mu \ud x^\nu \mathrm{,}
\eqx
where $u^\mu$  is the boost velocity parameter and $b$ is a dilatation
parameter related to the black brane temperature $T$ by $b = 1/\pi T$. 
The key ingredient of this approach is the introduction the incoming
Eddington-Finkelstein time coordinate. At the boundary this coordinate
reduces to the usual Minkowski time. Gauge/gravity duality 
maps physics at the boundary to the bulk of the AdS spacetime. For an
effective description at long wavelengths (relativistic hydrodynamics) the
map is provided by decreasing $r$ while keeping the Eddington-Finkelstein
time coordinate fixed \cite{Bhattacharyya:2008xc}. For
a Bjorken expansion it is natural to use proper time $\tau$ instead of the
usual Minkowski time, so an analogous Eddington-Finkelstein type proper
time coordinate $\ttau$ is introduced. 
Specifically, $u = \partial_{\tau}$ at the boundary, but is now taken as $u
= \partial_{\ttau}$ in the bulk of AdS space. Furthermore, for a boost
invariant flow the temperature $T$ is asymptotically proportional to
$\tau^{-1/3}$, which translates to $b = 3^{1/4} 2^{-1/2} \ttau^{1/3}$
(to ensure agreement with \cite{Janik:2005zt}). Thus finally 
\eq
\label{metricinf}
\ud s^{2} = - r^{2} (1 - \frac{4}{3 \ttau^{4/3} r^{4}}) \ud \ttau^{2} + 2
\ud \ttau \ud r + r^{2} \ttau^{2} \ud y^{2} + r^{2} \ud x_{\perp}^{2}
\mathrm{.}  
\eqx
It is straightforward to verify that metric \eqref{metricinf} is related to
the Janik-Peschanski metric  
\cite{Janik:2005zt}
\eq
\label{JP}
\ud s_{JP}^{2} = \frac{1}{z^{2}} \left\{ - \frac{\left( 1 - \frac{z^4}{3
    \tau^{4/3}} \right)^{2}} {1 + \frac{z^{4}}{3 \tau^{4/3}}} \ud\tau^2  +
\tau^{2} 
\left(1 + \frac{z^{4}}{3 \tau^{4/3}} \right) \ud y^{2} + \left(1 +
\frac{z^{4}}{3 \tau^{4/3}} \right) \ud x_{\perp}^{2} + \ud z^{2} \right\}
\eqx
by a coordinate transformation
\eqn
\label{coordtrafo}
\ttau &=& \tau \left\{ 1 - \frac{1}{\tau^{2/3}} \left[ \frac{3^{1/4}
    \pi}{4 \sqrt{2}} + \frac{3^{1/4}}{2 \sqrt{2}}
  \arctan{\left(\frac{3^{1/4}}{\sqrt{2}} r \cdot \tau^{1/3}\right)} +
  \frac{3^{1/4}}{4 \sqrt{2}} \log{\frac{r \cdot \tau^{1/3} -
      \frac{\sqrt{2}}{3^{1/4}}}{ r \cdot \tau^{1/3} +
      \frac{\sqrt{2}}{3^{1/4}}}} \right] \right\} \mathrm{,}
\nonumber \\ 
r &=& \frac{1}{z} \cdot \sqrt{1 + \frac{z^{4}}{3 \cdot \tau^{4/3}}}
\mathrm{.} 
\eqnx
Because of the nontrivial dependence of $\ttau$ on $r$ the
limits $\tau \rightarrow \infty$ and $\ttau \rightarrow \infty$
(corresponding to equilibration of gauge theory plasma) differ in the
bulk. Moreover the metric \eqref{metricinf} is regular and invertible up to
the black brane singularity, which is not the case with \eqref{JP}. These
observations are crucial for the present approach to work. Note also that
the relation between $\tau$ and $\ttau$ is singular when 
$z=3^{1/4} \tau^{1/3}$. This is precisely the locus where the singularities 
found in \cite{Heller:2007qt} were encountered. 

\spacer

The metric \eqref{metricinf} is not an exact solution of Einstein equations 
$R_{M N} + 4 G_{M N} = 0$ -- there are subleading corrections coming from
derivatives of the velocity $u$ and temperature $T$. They correspond to the gradient expansion of the 
boundary energy momentum tensor  \cite{Balasubramanian:1999re}. 
For a boost-invariant flow the energy-momentum tensor 
$<T_{\mu \nu}>$ is determined by the energy density
$\epsilon(\tau)$ \cite{Janik:2005zt}. In the gradient expansion each
covariant derivative $\nabla u$ is damped by
$\frac{1}{L \cdot T}$ (L is the characteristic length scale of a
perturbation and $T$ is the fluid temperature). Because boost-invariant 
flow is characterized by $T \sim \tau^{-1/3} + \ldots$ and moreover $\nabla u \sim
\tau^{-1}$, we expect the following expansion of the 
energy density $\epsilon(\tau)$ \cite{Baier:2007ix}
\eq 
\epsilon (\tau) = \frac{1}{\tau^{4/3}} \left\{1 - 2 \eta_{0}
\frac{1}{\tau^{2/3}} + \left( \frac{3}{2} \eta_{0}^2 - \frac{2}{3} \eta_{0}
\tau_{\Pi}^{0} + \frac{2}{3} \lambda_{1}^{0} \right) \frac{1}{\tau^{4/3}} +
\ldots \right\} \mathrm{,}
\eqx
where $\eta_{0}$ and $\tau_{\Pi}^{0}$ are the viscosity
and relaxation time coefficients, whereas $\lambda_{1}^{0}$ is a new
transport coefficient introduced in \cite{Baier:2007ix}. The above
expansion is written in terms of $\tau$, since the energy density is
defined at the boundary. 

\spacer

The apparent curvature singularities
  in AdS encountered in \cite{Heller:2007qt} appear at third order in the large $\tau$ (gradient)
expansion. It is difficult to check what happens for a general flow at this
order. However the situation is much simpler for a boost-invariant flow,
since all the symmetries can be imposed from the outset. This leads to
the following ansatz for the metric \footnote{This ansatz is adequate
    in the large proper time regime; in general the $\ud y^{2}$ term in the
    metric needs to be modified to ensure that the limit $b \rightarrow \infty$
    leads to an empty AdS space, as explained in
    \cite{Kinoshita:2008dq}. 
} 
\eq\label{metricansatz}
\ud s^{2} = G_{M N} \ud x^{M} \ud x^{N} = - r^{2} N(\ttau,r) \ud
\ttau^{2} + 2 \ud \ttau \ud r + r^{2} \ttau^{2} e^{b(\ttau,r)} \ud y^{2} +
r^{2} e^{c(\ttau,r)} \ud x_{\perp}^{2} \mathrm{.}
\eqx
Introducing the scaling variable $v = r \cdot \ttau^{1/3}$ in analogy with 
what is done in 
\cite{Janik:2005zt, Nakamura:2006ih, Janik:2006ft, Heller:2007qt, 
  Benincasa:2007tp, Buchel:2008xr, Buchel:2008qw} one obtains the natural 
expansion of the metric components in $\ttau^{-2/3}$ on the gravity
side
\eqn
N(\ttau,r) &=& A(v) \exp {\left(\sum_{k>0} a_k(v)\ttau^{-2k/3}\right)}
\mathrm{,} \nonumber \\
e^{b(\ttau,r)} &=& B(v) \exp {\left(\sum_{k>0} b_k(v)\ttau^{-2k/3}\right)}
\mathrm{,} \nonumber \\
e^{c(\ttau,r)} &=& C(v) \exp {\left(\sum_{k>0} c_k(v)\ttau^{-2k/3}\right)}
\mathrm{.} 
\eqnx
Note that the scaling variable $v$ introduced here is different from the
one used in \cite{Janik:2005zt, Nakamura:2006ih, Janik:2006ft,
  Heller:2007qt,Benincasa:2007tp, Buchel:2008xr, Buchel:2008qw}, and so the
scaling limit considered here must be regarded as different. 
To obtain a uniform expansion of the Einstein equations $E_{M N} \equiv
R_{M N} + 4 G_{M N} = 0$ one needs to rescale them (see
\cite{Janik:2006ft}) according to $\hat{E} = \left( 
\ttau^{2/3} E_{\ttau \ttau} , E_{\ttau r}, \frac{1}{\ttau^{2/3}} E_{r r},
\frac{1}{\ttau^{4/3}} E_{y y}, \ttau^{2/3} E_{x_{\perp} x_{\perp}}
\right)$. This leads to 
\eq
\label{EinsteineqnsS}
\hat{E}(\ttau,r) = \hat{E}_{0}(r \cdot \ttau^{1/3}) + \frac{1}{\ttau^{2/3}}
\hat{E}_{1}(r \cdot \ttau^{1/3}) + \frac{1}{\ttau^{4/3}} \hat{E}_{2}(r
\cdot \ttau^{1/3}) + \frac{1}{\ttau^{2}} \hat{E}_{3}(r \cdot \ttau^{1/3}) +
\ldots 
\eqx
The curvature invariants (e.g. $\mathcal{R}_{M N O P}
\mathcal{R}^{M N O P}$) defined recursively in \cite{Benincasa:2007tp} can 
be likewise expanded. The
crucial difference between the present approach and the one introduced in 
\cite{Janik:2005zt}  is that the expansion parameter involves $\ttau$
instead of 
$\tau$. Einstein equations can be solved order by order in $\ttau^{-2/3}$ 
expansion starting from
\eqn\label{orderzero}
A\left( v \right) &=&  1 - \frac{4}{3 v^{4}} \mathrm{,} \nonumber \\ 
B\left( v \right) &=& C\left( v \right) = 1 \mathrm{,}
\eqnx
which simply reproduces the boosted black brane solution
\eqref{metricinf}. 
Thus the zeroth order solution entails large but finite 
$\ttau$; it is to be expected that the singularity at $r=0$ should be
shielded by an event horizon, however it is difficult to demonstrate this
explicitly\footnote{Very recently it was shown in
    \cite{Kinoshita:2008dq} that an apparent horizon is present.}.

\section*{Gravity dual of the gradient expansion} 

The equations of motion \eqref{EinsteineqnsS} at a given order $k$ are a
system of ordinary second order differential equations for the 3 functions
$a_{k}\left( v \right)$, $b_{k} \left( v \right)$ and $c_{k} \left( v
\right)$. Each solution involves two integration constants. On the other
hand, two of the equations of motion are constraints. At each order $k>0$
one of the constraints fixes one of the integration constants appearing at
that order, and the other one fixes an integration constant left
undetermined at order $k-1$. The $4$ remaining integration constants can be
fixed order by order by imposing metric regularity (up to the usual black
brane singularity at $v = 0$ \cite{Bhattacharyya:2008jc}). It turns out
that the potential singularity is located only at $v = \sqrt{2}/3^{1/4}$;
thus the functions $b_k\left( v \right), c_k\left( v
\right)$ must remain finite as $v \rightarrow
\sqrt{2}/3^{1/4}$. In case of $a_k\left(v \right)$ the requirement should
be that the product with $A(v)$ must be finite. However one can take
advantage of residual diffeomorphism invariance \cite{Kinoshita:2008dq}
preserved by the ansatz \eqref{metricansatz}, which can effectively be
fixed by requiring that $a_{k}\left(v\right)$ itself be
regular. Furthermore asymptotic AdS behavior of the metric requires that
these functions vanish as $v \rightarrow \infty$ (in the late proper time
regime).  These 
conditions together with the constraints fix 5 of the 6 integration
constants at a given order $k>0$ and 
lead to a regular metric with no poles or logarithmic singularities apart
from $v = 0$. As an example, the first order solution (dual to viscous
hydrodynamics \cite{Nakamura:2006ih}) reads 
\eqn 
a_{1} &=& - \frac{2}{3}
\cdot \frac{2 \cdot 3^{-1/2} + 2^{1/2} 3^{-1/4} v + v^{2}} {\left( 2^{1/2}
  3^{-1/4} + v \right ) \cdot \left( 2 \cdot 3^{-1/2} + v^{2} \right)}
\mathrm{,} \nonumber \\ b_{1} &=& \frac{\pi}{\sqrt{2} 3^{3/4}} -
\frac{\sqrt{2}}{3^{3/4}} \arctan{\left(\frac{3^{1/4}}{\sqrt{2}} v \right)}
- \frac{2 \sqrt{2}}{3^{3/4}} \log{v} + \frac{\sqrt{2}}{3^{3/4}} \log{\left(
  \frac{\sqrt{2}}{3^{1/4}} + v \right)} + \frac{1}{\sqrt{2}3^{3/4}}
\log{\left( \frac{2}{3^{1/2}} + v^{2} \right)}  \mathrm{,} 
\eqnx 
and $c_{1}=-b_1/2$. 
Higher
order formulae (up to the third order) can be found in a Mathematica notebook available online
\cite{Math}. 
Holographic renormalization correctly reproduces the energy density for the
boost-invariant flow up to second order in derivatives \cite{Heller:2007qt,
  Baier:2007ix}.

\section*{Absence of singularities and relation to Fefferman-Graham
  coordinates} 

The assumption of non-singularity of coefficients of curvature invariants
in the late proper-time expansion was crucial in
establishing transport properties of $\mathcal{N} = 4$ SYM plasma in
\cite{Janik:2006ft,Heller:2007qt}. The present approach to boost-invariant
flow starts from a manifestly regular metric 
in the leading order (no logarithmic and power-like singularities at $v =
\sqrt{2}/3^{1/4}$) and produces regular solutions up to the third
order. 
Since the components of the metric as well as its inverse are regular
(as well as their derivatives), 
all curvature invariants are non-singular. 
Indeed, from \eqref{metricansatz} it 
follows that the non-vanishing components of the inverse are $G^{rr}= r^{2} 
e^{a(\ttau,r)}$, $G^{r\ttau}=1$, $G^{yy}= r^{-2} \ttau^{-2} e^{-b \left(
  \ttau, r \right)}$, 
$G^{\perp\perp}= r^{-2} e^{-c \left( \ttau, r \right)}$. 
If $e^{-a(\ttau,r)}$ had been present, singularities would have
appeared as a consequence of \eqref{orderzero}. 

\spacer

It is natural to ask how these results are related to those obtained using
the original approach of \cite{Janik:2005zt,Heller:2007qt}. Clearly, 
they should be related by a coordinate
transformation order by order in the large proper-time expansion:
\eqn
\ttau &=& \tau \Big(t_0 (\frac{z}{\tau^{1/3}}) + \frac{1}{\tau^{2/3}}
t_1(\frac{z}{\tau^{1/3}}) + \frac{1}{\tau^{4/3}} t_2(\frac{z}{\tau^{1/3}})
+ \frac{1}{\tau^{2}} t_3(\frac{z}{\tau^{1/3}})
+ \dots \Big)\\
r &=& \frac{1}{z} \Big(r_0(\frac{z}{\tau^{1/3}}) + \frac{1}{\tau^{2/3}}
r_1(\frac{z}{\tau^{1/3}}) + \frac{1}{\tau^{4/3}} r_2(\frac{z}{\tau^{1/3}})
+\frac{1}{\tau^{2}} r_3(\frac{z}{\tau^{1/3}})
+ \dots \Big)\ \ .
\eqnx
It is straightforward (though tedious) to determine this transformation
explicitly; the results up to third order are online \cite{Math}.
This transformation defines a map between the two approaches and explains
how the results obtained in \cite{Heller:2007qt} were 
correct despite the singularities which were encountered there. 

\section*{Summary}

This Letter shows how to construct a consistent gravity dual to 
boost-invariant flow of $\mathcal{N} = 4$ SYM plasma in a gradient
expansion. This solution reproduces known results up to second order
viscous hydrodynamics \cite{Heller:2007qt}.  A manifestly regular metric up
to third order 
was described and arguments were given why all curvature invariants have
non-singular expansions in $\ttau^{-2/3}$. The relation with the approach
based on Fefferman-Graham coordinates was made explicit by the coordinate
transformation discussed above.

\spacer

In conclusion, the AdS/CFT dual to boost-invariant flow can be realized
within the supergravity approximation and is completely free of
singularities apart from the black brane singularity at $r=0$.

\spacer 

\noindent {\em Note added: } Shortly after the original version of this
Letter was posted to {\tt arXiv.org} a very interesting paper
\cite{Kinoshita:2008dq} appeared which deals with the same subject and
along similar lines as the approach disussed here. Apart from demonstrating
the presence of an apparent horizon, the authors also give an all-orders
argument for the absence of singularities in the large proper time
expansion.

\begin{acknowledgments}
MH thanks the following institutions for their hospitality and financial
support: the Institute for Nuclear Theory at the University of Washington,
So{\l}tan Institute for Nuclear Studies in Warsaw, Durham University and
Perimeter Institute. Research at Perimeter Institute is supported by the
Government of Canada through Industry Canada and by the Province of Ontario
through the Ministry of Research \& Innovation. PS thanks the Tata
Institute of Fundamental Research and the Galileo Galilei Institute for
Theoretical Physics for its hospitality and support during the completion
of this work. MH and PS were supported by Polish Ministry of Science and
Information Technologies grant 1P03B04029 (2005-2008) during initial stages
of this work. While finishing this project MH was supported by Polish
Ministry of Science and Information Technologies grant N N202 247135
(2008-2010) and the British Council Young Scientists Programme scholarship.

\spacer

We acknowledge the use of Matthew Headrick's excellent
Mathematica package {\tt diffgeo.m} \cite{Headrick}. 

\spacer


The authors would like to thank Shiraz Minwalla for discussions and Alex
Buchel, Veronica Hubeny, Romuald Janik, Shiraz Minwalla, Robi Peschanski
and Mukund Rangamani for reading the manuscript and helpful comments.

\end{acknowledgments}

\end{document}